\documentclass[conference]{IEEEtran}
\usepackage{enumerate}

\ifCLASSOPTIONcompsoc
  \usepackage[nocompress]{cite}
\else
  \usepackage{cite}
\fi

\ifCLASSINFOpdf

\usepackage[pdftex]{graphicx}

\DeclareGraphicsExtensions{.pdf,.jpeg,.png}

\else
\fi

\usepackage{amsmath}

\usepackage[hyphens]{url}

\usepackage{fancyhdr}  

\begin{document}
%

\title{Verifiable Anonymous Identities and Access Control in Permissioned Blockchains}


\author{
\IEEEauthorblockN{Thomas Hardjono}
\IEEEauthorblockA{MIT Internet Trust Consortium\\
Massachusetts Institute of Technology\\
Cambridge, MA 02139, USA\\
Email: hardjono@mit.edu}
\and
\IEEEauthorblockN{Alex (Sandy) Pentland}
\IEEEauthorblockA{MIT Connection Science \& MIT Media Lab\\
Massachusetts Institute of Technology\\
Cambridge, MA 02139, USA
\\Email: sandy@media.mit.edu}
}


%


\maketitle

\pagestyle{fancy} 
\thispagestyle{fancy}  
\lhead{DRAFT} 
\rhead{April 17, 2016}

\begin{abstract}
In this paper we address the issue of identity and access control
within shared permissioned blockchains.
We propose the ChainAchor system that provides
{\em anonymous but verifiable} identities for entities on the blockchain.
ChainAchor also provides {\em access control} to entities seeking
to submit transactions to the blockchain to read/verify transactions
on the the permissioned blockchain.
Consensus nodes enforce access control to the shared permissioned blockchain
by a simple look-up to a (read-only) list of anonymous members' public-keys.
ChainAnchor also provides {\em unlinkability} of transactions belonging to an entity on the blockchain.
This allows for an entity to optionally disclose their identity
when a transaction is called into question (e.g. regulatory or compliance requirements),
but without affecting the anonymity and unlinkability of their remaining transactions.
~~\\


\end{abstract}



%
\IEEEpeerreviewmaketitle

\section{Introduction}
\label{sec:intro}

The recent rise to prominence of the Bitcoin \cite{bitcoin} decentralized digital currency
system has generated broad interest
in blockchains as a new form of infrastructure for maintaining a shared 
and cryptographically immutable ledger.
Consequently interest has also peaked in the possible development of
{\em permissioned} and {\em private} blockchain systems,
in contrast to the
{\em permissionless} and {\em public} blockchain in Bitcoin.

There are numerous use-cases for permissioned blockchain systems
whose goal is to provide an immutable ledger that
captures the existence of digital facts or artifacts (e.g. transactions, documents, etc) 
in a given moment in time in a non-repudiable manner.
In order to understand the differences among these private permissioned blockchain systems,
we believe it is useful to further distinguish between closed and shared permissioned blockchain systems.
We define a {\em closed} permissioned blockchain as one in which the entities having access to the private blockchain
all belong to the same organization having a common business interest.
We define a {\em shared} permissioned blockchain as a private blockchain where the transacting entities
belong to distinct organizations with competing interests.

In looking at closed and shared permissioned blockchain systems
there a number of challenges with regards to identity and access control to the blockchain:
\begin{itemize}

\item	{\em Identity privacy}:
The issue of identity privacy can be acute when a blockchain is
shared among competing entities.
There is a potential that the behavior of an identity on a blockchain
may inadvertently disclose or leak information to competitors transacting on the same blockchain.

\item	{\em Access control}:
In a shared permissioned blockchain, controlling access (both read-access and write-access)
is crucial to the value of the blockchain.
New approaches are needed beyond the classic Enterprise access control regime
that prioritize the security of the infrastructure and services
over the identity privacy of entities using the infrastructure.

\item	{\em Optional disclosure \& transaction privacy}:
New approaches are needed to provide transaction privacy in the form
of {\em unlinkability} of transactions.
We believe that new solutions are needed for {\em optional disclosure}
of identities relating to transactions that come into questions (e.g. AML or regulatory compliance).
This feature allows an individual to own multiple unlinkable transaction keys on the blockchain,
and disclose ownership of a key (e.g. upon legal challenge)
without affecting the security and privacy of his or her remaining other keys on
the same shared permissioned blockchain.

\end{itemize}

In this paper we propose the ChainAnchor system that addresses these challenges.
The current work expands and generalizes our previous work~\cite{HardjonoSmith2016a}
that addressed constrained devices in an IoT environment.
In the next section we describe the ChainAnchor architecture and protocol steps.
The design of ChainAchor aims to be functionally independent from the
underlying blockchain system, with the goal of deployability
of ChainAnchor on various blockchains.

The current paper seeks to be readable to a broad audience,
and as such it does not
cover in-depth the cryptography behind EPID~\cite{BrickellLi2012} and DAA~\cite{Brickell2004}
schemes that provide for identity anonymity.
We assume the reader is familiar with public-key cryptography
and with the basic operations of the blockchain in the Bitcoin system.
In order to assist the more curious reader, we provide 
a brief summary of the EPID scheme in the Appendix
and provide pointers to the relevant equations in the Appendix.
The current paper focuses on an RSA-based EPID scheme
based on the Camenisch-Lysyanskaya signature scheme~\cite{CamenischLysyanskaya2002}
and the DAA scheme of Brickell, Camenisch and Chen~\cite{Brickell2004}.
Readers are directed to the authoritative papers
of~\cite{Brickell2004} and~\cite{BrickellLi2012} for an in-depth discussion.
An EPID scheme using bilinear pairings can be found in~\cite{Brickell2010Bilinear}.
It is based on the Boneh, Boyen and Schacham group signature scheme~\cite{BonehBoyenShacham2004}
and the Boneh-Schacham group signature scheme~\cite{BonehShacham2004}.

\begin{figure*}[!t]
\centering
\includegraphics[width=7in]{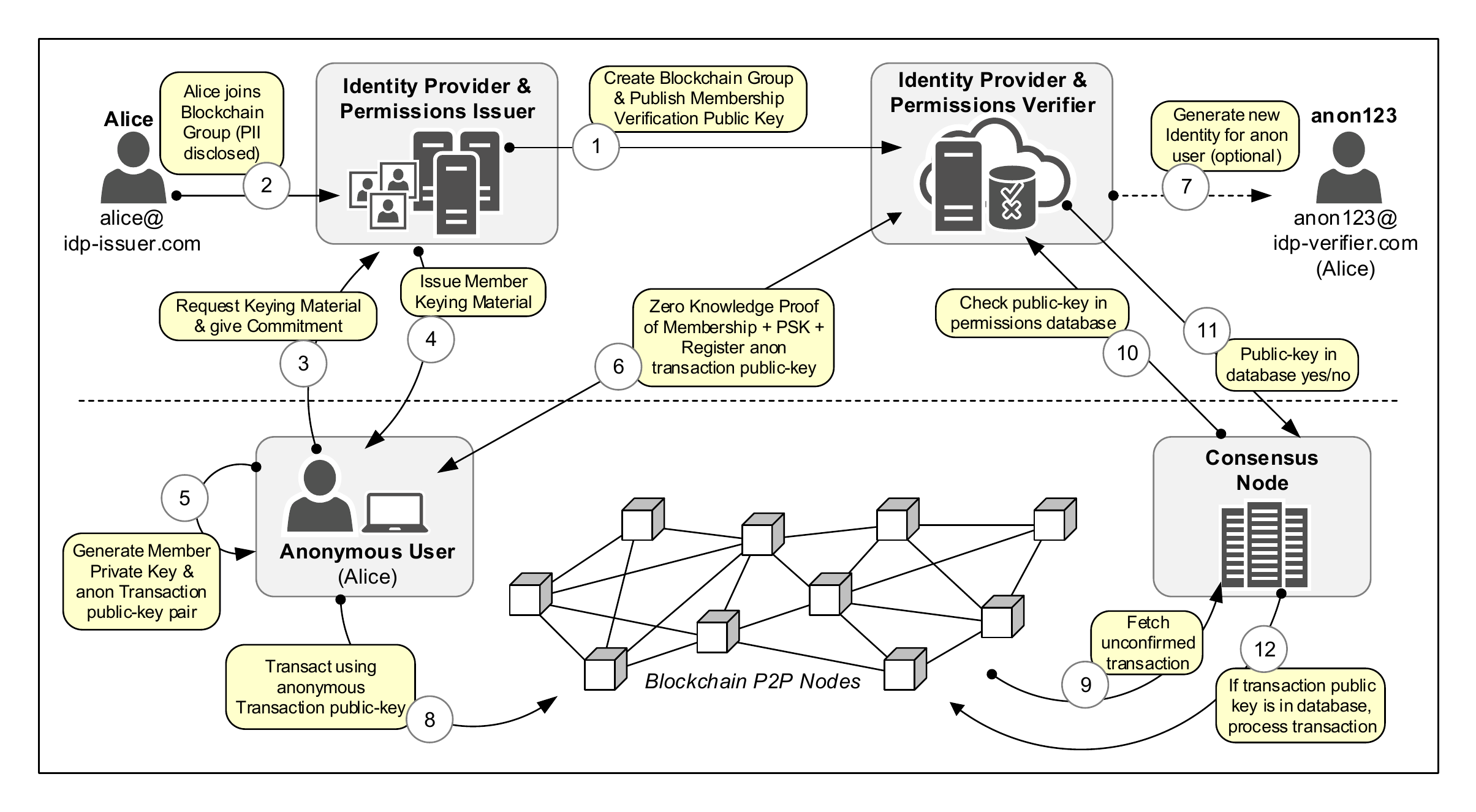}
\caption{Overview of ChainAnchor interactions}
\label{fig:chainanchoroverview}
\end{figure*}

\section{ChainAnchor Architecture}
\label{sec:chainanchor-protocol}

As mentioned previously,
we seek to address two important aspects of blockchain systems,
namely that of the privacy of the identity of the user and
that of providing access control to a permissioned blockchain.
Figure~\ref{fig:chainanchoroverview} summarizes the ChainAnchor entities and interactions.

Our proposed ChainAnchor system generalizes
our previous device-centric solution reported in~\cite{HardjonoSmith2016a},
and provides the following features:
\begin{itemize}
\item	 {\em Anonymous but verifiable identities}:
Firstly, ChainAnchor allows participants in a shared permissioned blockchain
to maintain their privacy by allowing them to use
a {\em verifiable anonymous identity} when transacting on the blockchain.
The anonymity (pseudonymity) of their identity is achieved
using the EPID zero-knowledge proof scheme~\cite{BrickellLi2012},
which allows the owner to remain truly anonymous in transactions
but allows other parties to verify that the identity is genuine.
The verification process is such that no PII or other identifying information disclosed
about the owner of the identity.
~~\\

\item	{\em Access control to blockchain}:
Secondly, the anonymous (but verifiable) identities make-up the ``group''
of entities allowed to access the permissioned blockchain.
An entity who has proved their membership using the zero knowledge proof (ZKP) protocol
can then ``register'' their self-asserted transaction
public-key (for transacting on the blockchain).
The consensus nodes collectively enforce access control
to the blockchain by only processing transactions from the members'
transaction public-keys.
Transactions originated from (destined to) unknown public-keys are simply ignored or dropped.
~~\\

\item	{\em Multiple unlinkable transaction keys}:
A member of a shared blockchain can execute the 
ZKP protocol as many times as they desire,
each time registering a different self-asserted transaction public-key.
No one on the blockchain (including the owner of the blockchain)
has the ability to link these keys to the single owner.
This allows a member to optionally disclose their identity
(e.g. when challenged in a regulatory context)
without endangering their other transactions public-keys.
~~\\

\end{itemize}

When a user requests membership to the permissioned blockchain,
the user sends this request to the a {\em Permissions Issuer} entity
who implements the permissioned group on behalf
of the owner of the permissioned blockchain.
For simplicity, we assume that the group owner
and the owner of the permissioned blockchain are the same entity.
In this step, the real-world identity of the user
will be known to the group owner and Permissions Issuer entity.
The user could  represent himself or herself,
or represent an organization who is a member of the group
sharing the permissioned blockchain.
The precise credentials and attributes
required for membership approval is outside the scope of the current work.

Once the user is approved to join the permissioned blockchain
by the group owner,
the Permissions Issuer provides the user
with a number of user-specific keying material.
This keying material is crucial for the user to
later prove membership and to then register the public-key
intended to be used on the permissiond blockchain
-- referred to as the user's {\em transaction public key}.

The entity to whom the user must prove membership in an anonymous fashion
and register the user's transaction public key
is the {\em Permissions Verifier} entity.
This entity must be distinct from the Permissions Issuer entity
in order to protect the privacy of the user.
Note that since the user's transaction public-key pair is self-generated,
only the user knows the transaction private-key.

The user proves membership to the Permissions Verifier
by using the zero-knowledge proof ZKP protocol,
which by design protects the anonymity of the user.
Once a user successfully completes this protocol, 
the user's transaction public-key is delivered by the user to the Permissions Verifier
under a secure channel.
In turn the Permissions Verifier will add the user's transaction public-key
to a {\em Permissions Database} for that group.

This permissions database -- which can be a simple list --
 is maintained by the Permissions Verifier entity
and is read-accessible by the consensus nodes (i.e. ``miners'') in the permissioned blockchain.
A user can have as many transaction public-keys as they wish
on the same blockchain. This is achieved by the user
executing a distinct run of the zero-knowledge proof protocol,
each time registering a different transaction public-key.

The database only holds
the transaction public-keys of the members
and the timestamp of the completion of the zero-knowledge proof protocol execution.
The Permissions Verifier is not able to distinguish one user from another.
Furthermore, the Permissions Verifier is not able to know
whether or not a user has multiple transaction public-keys 
registered in the permissions database.

Access-control for the permissioned blockchain is enforced by the consensus nodes
in the blockchain, based on the list of (anonymous) public-keys in the permissions database.
Prior to processing an unconfirmed transaction,
a consensus node enforcing access-control must first
verify that the public-key associated with the transaction
is present in the permissions database for that blockchain.
In other words, the consensus nodes in the permissioned blockchain
must ensure that the blockchain contains
transactions only from anonymous users whose transaction public-keys
are listed in the permissions database.

In this paper we propose the functions of the Permissions Issuer
and Permissions Verifier to be implemented
by distinct identity provider (IdP) entities.
This is because the IdP is the traditional issuer
of a digital identities on the Internet
and thus possess the necessary infrastructure
implementing standard protocols
relating to identity management and identity federation.
Furthermore, many Enterprise organizations today are already
operating an IdP service (or server) internally.
As such, an Identity Federation (e.g. via SAML2.0 or OpenID-Connect)
can be established for 
organizations which are sharing a permissioned blockchain.

\subsection{Entities in the System}

\begin{itemize}

\item	{\em Identity Provider and Permissions Issuer} (IdP-PI):\\
The IdP-PI is the identity provider entity that establishes the permissioned group
on behalf of the group owner.
For a given permissioned group, there is exactly one IdP-PI entity.
~~\\

\item	{\em Permissions Verifier} (IdP-PV):\\
The IdP-PV is the identity provider entity that performs
the anonymous group-membership verification of a given a User
by running the zero-knowledge proof protocol with that User.
The IdP-PV maintains the Permissions Database.
For a given permissioned group, there can be multiple independent
IdP-PV entities (although only one IdP-PI for the group).
~~\\

\item	{\em User}:\\
The User is the entity wishing to join the permissioned blockchain
(i.e. group permissioned)
implemetted by the IdP-PI and the IdP-PV.
~~\\

\item	{\em Consensus node}:\\
The Consensus Node (``miner'') is entity processing transactions
from valid group members of a permissioned blockchain.
~~\\

\item	{\em Owner}:\\
Although not shown explicitly in Figure~\ref{fig:chainanchoroverview},
a permissioned group must be owned
by an organization or individual.
We use the term {\em Owner} for this entity.

\end{itemize}

\subsection{Keys in the System}
\label{sebsec:keys-in-system}

A major feature of the EPID zero-knowledge proof scheme~\cite{BrickellLi2012} 
is its ``signature of proof'' mechanism,
which allows multiple distinct private-keys to be used
with one public-key.
This allows each distinct User to deploy individually
unique EPID private-keys (which they keep secret),
from which any signature can be verified by the IdP-PV using the single EPID public-key.
We refer to this public-key as the {\em membership verification public-key},
and the multiple distinct private-keys
as the {\em user-member private-key}.

More specifically, for a given permissioned group ${{PG}}$ there is one (1)
membership verification public key $K_{\operatorname{PG}}$ held by the IdP-PV entity.
That one public-key is used by the IdP-PV
to validate the membership of multiple ($n$) users
$U_1, \ldots, U_n$ whose corresponding user-member private keys are
$K_{\operatorname{PG-{U_1}}}^{-1}, \dots, K_{\operatorname{PG-{U_n}}}^{-1}$.

The keys in the ChainAnchor system are summarized below.
We adopt the notational convention of~\cite{BrickellLi2012}
by denoting a public-key pair as $(K, K^{-1})$, with the public-key being $K$.
\begin{itemize}

\item	{\em Membership Issuing Private Key}:\\
This key is denoted as $K_{MIPK}^{-1}$ and is
is generated by the IdP-PI for each permissioned group that the
IdP-PI establishes. 
This key is unique for each permissioned group.
This key is used by the IdP-PI
in enrolling or adding new Users to the permissioned group.
(This key is shown in {Eq.}~\ref{eq:group-issuing-private-key}
in the Appendix).
~~\\

\item	{\em Membership Verification Public Key}:\\
This key is denoted as $K_{\operatorname{PG}}$ and is
generated by the IdP-PI and is delivered over a secure channel
to the Permissions Verifier entity (IdP-PV).
This key is unique for each permissioned group.
(This key is shown in {Eq.}~\ref{eq:group-public-key}
in the Appendix).
~~\\

\item	{\em User-Member Private Key}:\\
For a given User $U_i$ who is a member in a permissioned group ${PG}$,
the user-member private key is denoted as 
$K_{\operatorname{PG-{U_i}}}^{-1}$.
(This key is shown in {Eq.}~\ref{eq:membership-private-key}
in the Appendix).
~~\\

\item	{\em User's Transaction Public-Key Pair}:\\
This is the transaction public-key pair
that the User employs to transact.
This key pair is generated by the User
(i.e. user's computer or device).
We denote this public-key pair
as $(K_{trans}, K_{trans}^{-1})$,
with the public key being $K_{trans}$.
~~\\

\item	{\em User's Identity Public-Key Pair}:\\
This is the public-key pair
that the User employs to represent himself/herself
as a member of the permissioned-group.
This key pair is generated by the User
(i.e. user's computer or device).
We denote this public-key pair
as $(K_{id}, K_{id}^{-1})$,
with the public key being $K_{id}$.
~~\\

\item	{\em User \& IdP-PV Pairwise Shared Key (PSK)}:\\
As part of proving group-membership, the User and the 
DB-PV will establish a pairwise shared key (PSK).
The PSK is a symmetric key.
~~\\

\item	{\em IdP-PI and IdP-PV Certificates}:\\
These are traditional public-key pairs and X509 certificates:
\begin{itemize}
\item	{\em IdP-PI public-key pair}:
We denote the public key pair of the IdP-PI as
$(K_{PI}, K_{PI}^{-1})$ with the public key being $K_{PI}$.
~~\\

\item	{\em IdP-PV public-key pair}:
Similarly, we denote the public key pair of the IdP-PV as
$(K_{PV}, K_{PV}^{-1})$ with the public key being $K_{PV}$.
~~\\

\end{itemize}

\end{itemize} 

\subsection{ChainAnchor Protocol Steps}
\label{subsec:ChainAnchorProtocolSteps}

In the following, we describe the steps of the ChainAnchor
design (see Figure~\ref{fig:chainanchoroverview}).
\\~~

\noindent {\bf{[Step~0]}} {\em IdP-PI Establishes Permissioned Group}:

This step is not shown in Figure~\ref{fig:chainanchoroverview}.
As part of the creation of a permissioned group,
the IdP-PI generates a number parameters that are unique
to the permissioned group and are used
to create two important keys related to the function
of the IdP-PI as the Permissions Issuer:
\begin{itemize}

\item	{\em Membership Verification Public Key}: $K_{\operatorname{PG}}$\\
The IdP-PI creates this key
to be used later by the Permissions Verifier entity (IdP-PV) when
engaging the User in the zero-knowledge proofs protocol.
(See Equation~\ref{eq:group-public-key} in Appendix A).
~~\\

\item	{\em Membership Issuing Private Key}: $K_{MIPK}$\\
The IdP-PI creates this key in order to issue unique keys
to Users in the system that allows the User later to prove 
membership to the IdP-PV.
(See Equation~\ref{eq:group-issuing-private-key} in Appendix A).
This issuing private key is kept secret by the IdP-PI.

\end{itemize}
~~\\

\noindent {\bf{[Step~1]}} {\bf {\em IdP-PI Shares Verification Public Key with IdP-PV}}:

In this step, the IdP-PI makes known the Membership Verification Public Key ($K_{\operatorname{PG}}$)
to the IdP-PV.
We assume a secure channel with mutual authentication is used
between the IdP-PI and IdP-PV entities.

~~\\

\noindent {\bf{[Step~2]}} {\bf {\em User Authenticates \& Requests Membership}}

To join the permissioned group
the User  sends the request to the IdP-PI
that manages the permissioned group of interest.
The User must first authenticate itself to the IdP-PI.
The method used to authenticate is outside the scope of the current paper.

At this point in the ChainAnchor protocol the User is not anonymous to the IdP-PI,
and it knows the true identity of the User
(e.g. has an account such as {\tt alice@idp-issuer.com} at the IdP-PI).

A User who has successfully authenticated and obtained approval
to join the group is then given a copy
of the Membership Verification Public Key $K_{\operatorname{PG}}$
by the IdP-IP using a secure channel.
~~\\

\noindent {\bf{[Step~3]}} {\bf {\em User Delivers Blinded Commitment Parameters}}

After obtaining Membership Verification Public Key
$K_{\operatorname{PG}}$ for the relevant group,
the User perform the following tasks:
\begin{itemize}

\item	{\em User validates the Membership Verification Public Key}:
Prior to using $K_{\operatorname{PG}}$ the User
must verify that the components in $K_{\operatorname{PG}}$ are formed correctly
(see Equation~\ref{eq:group-public-key} in Appendix A).
~~\\

\item	{\em User generates blinded commitment parameters}:
The User employs some of the parameters in $K_{\operatorname{PG}}$
to create his/her own {\em commitment} parameters that ``blinds'' 
the User's own secret keying material to the IdP-PI.
(See Equations~\ref{eq:user-private-U} and~\ref{eq:user-private-f} in Appendix A).
~~\\

\item	{\em User sends blinded commitment parameters to the IdP-PI}:
The User sends the commitment parameters to the IdP-PI,
who in-turn must verify that these parameters are formed correctly.
~~\\

\end{itemize}

It is important to note here that
the cryptographic {\em blinding} (in the commitment values) 
is done to retain the anonymity of the User to the IdP-PI.
The IdP-PI must unable to distinguish
one user from another at this point
based on a user's blinded commitment parameters.
~~\\

\begin{figure*}[!t]
\centering
\includegraphics[width=5in]{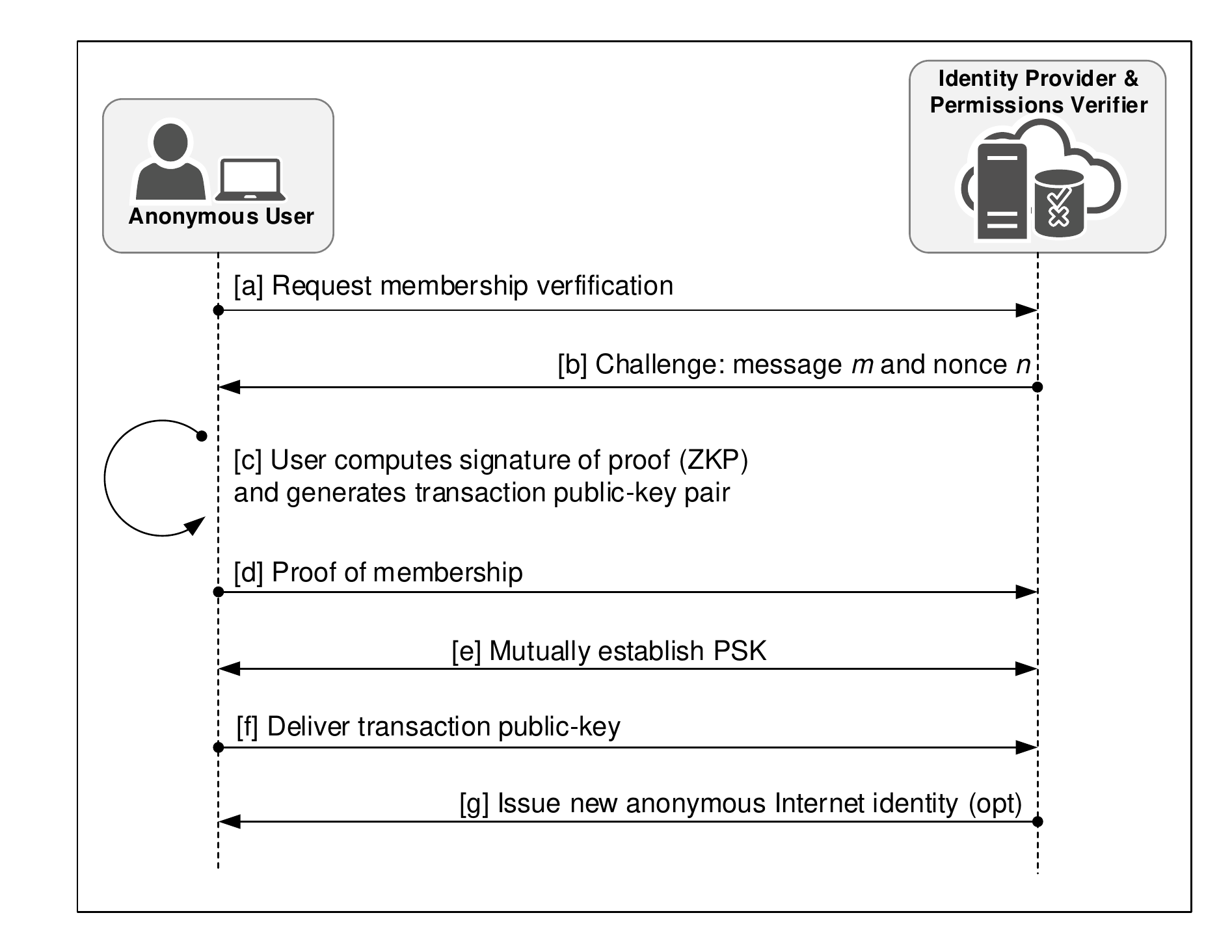}
\caption{User Anonymously Proves Membership to IdP-PV}
\label{fig:chainanchor-proving-membership}
\end{figure*}

\noindent {\bf{[Step~4]}} {\bf {\em IdP-PI Responds with Group-Member Keying Parameters}}

In this step, after validating
the blinded commitment parameters the IdP-PI 
generates a number of parameters associated with the
Member Private Key and sends them to the User.
(This is shown in
{Eq.}~\ref{eq:issuer-keying-material} in the Appendix).
In-turn the User uses these parameters to generate its own 
user-member private key (in the next step).
~~\\

\noindent {\bf{[Step~5]}} {\bf {\em User Generates User-Member Private Key}}

Upon receiving the user-specific group-member keying parameters,
the User uses these parameters to generate his/her
own {\em User-Member Private Key}, denoted as $K_{\operatorname{PG-{U_i}}}^{-1}$
(See Equation~\ref{eq:membership-private-key} in Appendix A).
Additionally, the User also generates its blockchain transaction key pair
$(K_{Trans}, K_{Trans}^{-1})$.
~~\\

\noindent {\bf{[Step~6]}} {\bf {\em User Proves Membership to IdP-PV}}

The anonymous membership verification protocol consists of a number of sub-steps
following the challenge-response model.
The User sends a request to the IdP-PV,
and in-turn the IdP-PV challenges the User with some parameters
that the User must respond to.

The sub-steps of the anonymous membership verification protocol are as follows
(Figure~\ref{fig:chainanchor-proving-membership}):

\begin{itemize}

\item	{\bf Step~6.1}:  The User sends a request to 
the IdP-PV for an anonymous membership verification
(Figure~\ref{fig:chainanchor-proving-membership}(a)).
~~\\

\item	{\bf Step~6.2}:  The Permissions Verifier IdP-PV responds by
returning a challenge message $m$ and a random nonce $n_{pv}$ to the User
(Figure~\ref{fig:chainanchor-proving-membership}(b)).
~~\\

\item	{\bf Step~6.3}:  Upon receiving the challenge message $m$ and
the random nonce $n_{pv}$ from the verifier IdP-PV,
the User must compute a ``signature of knowledge'' 
of the commitment parameter that the User supplied to the IdP-PI in Step~2.
(See Figure~\ref{fig:chainanchor-proving-membership}(c)).
The signature-of-knowledge is denoted as $\sigma$.
(See Equation~\ref{eq:three-sigma} in Appendix A).

~~As input into the signature-of-knowledge  $\sigma$ computation,
the User inputs:
	\begin{itemize}

	\item	The Membership Verification 
	Public Key ($K_{\operatorname{PG}}$) for the group which the User obtained
	from the Permissions Issuer IdP-PI in Step~2. 
	(See Equation~\ref{eq:group-public-key} in Appendix A).

	\item	The User's own User-Member Private Key 
	$K_{\operatorname{PG-{U_i}}}^{-1}$ which the User computed
	in Step~3.
	(See Equation~\ref{eq:membership-private-key} in Appendix A).

	\item	The challenge $m$ and the nonce $n_{pv}$ obtained from the 
	Permissions Verifier IdP-PV.
	~~\\
	\end{itemize}

\item	{\bf Step~6.4}:  The User sends the computed
signature-of-knowledge value $\sigma$ to the IdP-PV
as proof of the user's membership in the group
(Figure~\ref{fig:chainanchor-proving-membership}(d)).
~~\\

\item	{\bf Step~6.5}:  The IdP-PV validates signature-of-knowledge $\sigma$, and
returns an acknowledgement of a successful verification process to the User
together with some parameters to establish a pair-wise shared key (PSK) between the User and the IdP-PV.
~~\\

\item	{\bf Step~6.6}:  The User and the IdP-PV engage in a key agreement subprotocol that results
in a pair-wise shared key (PSK) -- (Figure~\ref{fig:chainanchor-proving-membership}(e)).
This PSK is shared between the User (who is anonymous throughout Step~6)
and the IdP-PV.
~~\\

\item	{\bf Step~6.7}:   The User delivers his or her
transaction public-key $K_{trans}$ to the IDP-PV under
a secure channel created using the shared PSK.
The IdP-PV then adds the User's transaction public key $K_{trans}$ to the permissions database.
(Figure~\ref{fig:chainanchor-proving-membership}(f)).
~~\\

\end{itemize}

\noindent {\bf{[Step~7]}} {\bf {\em IdP-PV Creates Anonymous Internet Identity}}

Optionally, as the result of a successful anonymous membership verification of the User
in the previous step, the IdP-PV creates a new anonymous Internet identity for the user
(e.g. {\tt anon123@idp-verifier.com}).
The IdP-PV returns a copy of this new Internet identity
to the User (Figure~\ref{fig:chainanchor-proving-membership}(g))
under a secure channel created using the PSK.
~~\\

\noindent {\bf{[Step~8]}} {\bf {\em User Transacts using Transaction Key-Pair}}

In this phase the User transacts on the blockchain in the usual manner
using the transaction private-key $K_{trans}^{-1}$ to sign transactions.
~~\\

\noindent {\bf{[Step~9]}} {\bf {\em Consensus Receives or Fetches Transaction}}

The consensus node fetches a transaction (i.e. from the pool of unprocessed transactions)
and prepares to process that transaction.
~~\\

\noindent {\bf{[Step~10]}} {\bf {\em Consensus Node Validates User's Public Key}}

Prior to processing a transaction,
a consensus node participating in the ChainAnchor permissioned-group
must check that the public-key found in the transaction
has been approved to participate in the permissioned-group.
That is, the consensus node must first look-up the permissions databases
at the IdP-PV to ensure the public key is in the database.
~~\\

\noindent {\bf{[Step~11 \& 12]}} {\bf {\em Consensus Node Processes Transaction}}

If the public key $K_{trans}$ used in the User's transaction
exists in the permissions database at the IdP-PV,
the  consensus node can proceed with processing the transaction.
Otherwise the  consensus node can choose to ignore the transaction.
~~\\

\section{Verifiable Anonymous Identities}
\label{sec:VerifAnonIdentities}

The identities in ChainAnchor -- in the form of transaction public-keys --
are anonymous because the transaction public-key pairs are self-generated by the user
(just as in the Bitcoin systems and in the PGP system).
Only the user knows the private key(s).

These identities are verifiable because the user at any time can execute the zero knowledge proof protocol
with the IdP-PV in order to prove membership in the given blockchain.
The unlikability of the multiple public-key pairs (belonging to a single user)
is also derived from the zero knowledge proof protocol~\cite{BrickellLi2012}.

\begin{itemize}

\item	{\em User remain anonymous to IdP-PV}:
The IdP-PV cannot distinguish among validated users.
More specifically, if two Users $U_1$ and $U_2$ independently
returns the challenge message $m$
with a signature-of-knowledge (see Equation~\ref{eq:three-sigma})
created using keys $K_{\operatorname{PG-{U_1}}}^{-1}$ and
$K_{\operatorname{PG-{U_2}}}^{-1}$
then the IdP-PV can verify both signature
using the one verification public key $K_{\operatorname{PG}}$ but
it will not be able to distinguish between Users $U_1$ and $U_2$.
~~\\

\item	{\em User remain anonymous to IdP-PI}:
When the User requests membership to the group,
the User (person) is known to the IdP-PI.
However, after Step~{5} the User becomes anonymous even to the
IdP-PI because the User injects a secret ``blinding'' parameter (in Step~3)
when generating the User's {\em User-Member Private Key}
(see Equations~\ref{eq:user-private-U},~\ref{eq:user-private-f} and~\ref{eq:membership-private-key} in the Appendix).
Since the IdP-PI is not involved in Step~5 onwards,
the IdP-PI has no knowledge of which transaction public-key pairs are owned by the User.
~~\\

\item	{\em Optional Disclosure of Transaction Keys}:
A User can deploy as many transaction keys as they wish within the blockchain.
This is achieved by the User executing the ZKP protocol
with the IdP-PV for each of the keys he or she wishes to register and use in the blockchain.
This approach has the advantage that the User may reveal (to the IdP-PV) 
the User's ownership of a given transaction key
without affecting other transaction keys (i.e. unlinkability).
~~\\

\end{itemize}

\section{Access Control to the Shared Permissioned Blockchain}
\label{sec:VerifAnonIdentities}

\begin{figure}[!t]
\centering
\includegraphics[width=3.5in]{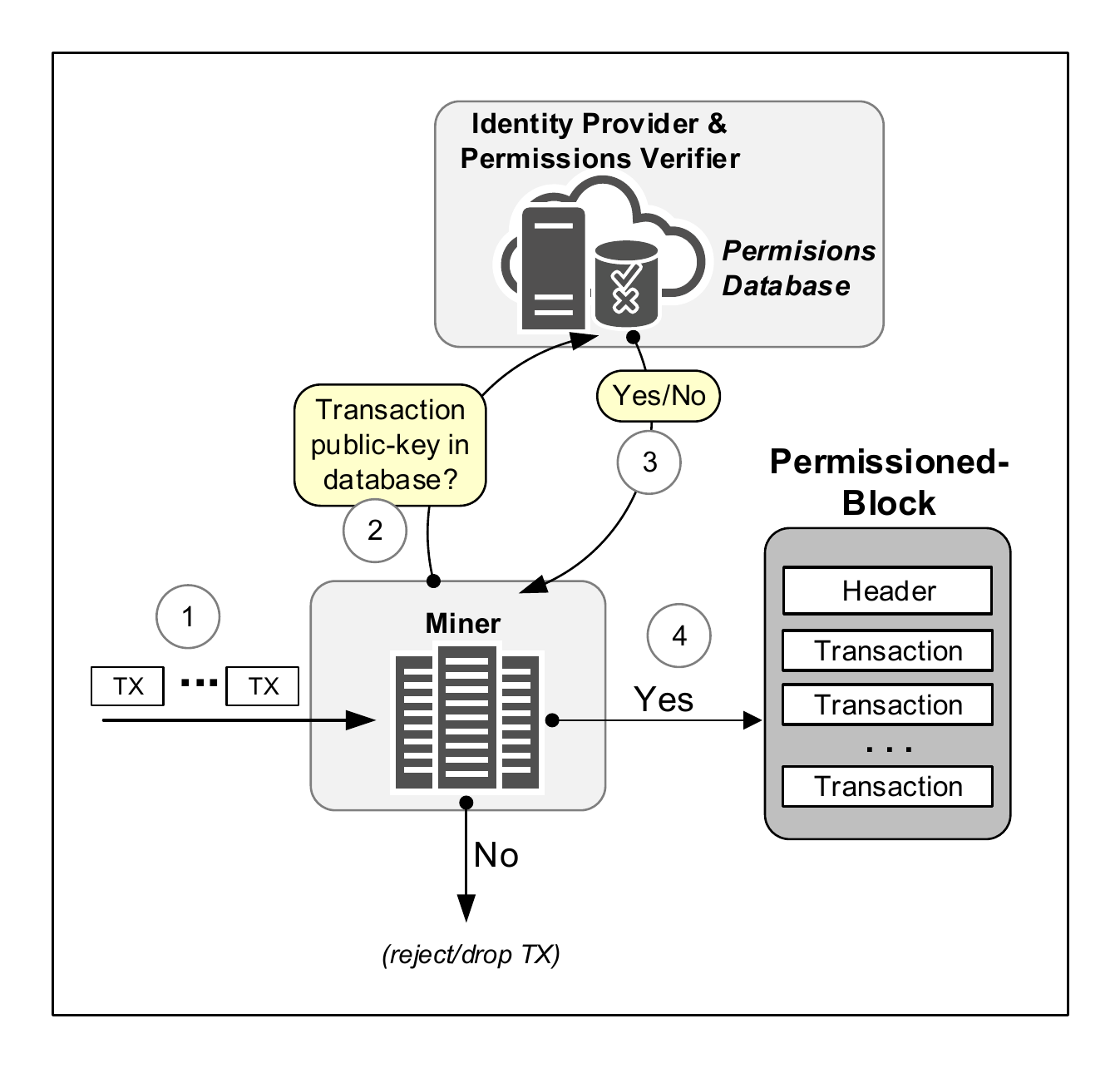}
\caption{Consensus node enforcing access control}
\label{fig:chainanchor-accesscontrol}
\end{figure}

In ChainAnchor the term ``consensus'' includes the notion of {\em membership} to a permissioned (private) blockchain.
ChainAnchor employs the consensus nodes (miners) in the shared permissioned blockchain
to collectively enforce access control,
by way of a simple ``filtering'' of transaction belonging to members
(see Figure~\ref{fig:chainanchor-accesscontrol}).
A consensus node needs to verify that the public-key of a transaction
that it wishes to process belongs to a member of the permissioned blockchain.
It does this verification by a simple look-up to the permissions database at IdP-PV.
(Other list-sharing methods can also be deployed, such as the node 
maintaining a local copy of the permissions database).

However, in order to counter possible cheating from dishonest nodes,
(who may allow non-member transaction to come into the blockchain)
ChainAnchor also requires that the IdP-PV and IdP-PI be {\em Validator} nodes
that checks that transactions belong to members of the blockchain.
The issue of the {\em weight} of a decision (vote) coming from the IdP-PV or IdP-PI
(versus coming from a consensus node) is outside the scope of the current paper
and will be the subject of further study.

\section{Anonymous Identities: Beyond the Blockchain}
\label{sec:AddressableIdentities}

\begin{figure}[!t]
\centering
\includegraphics[width=3.5in]{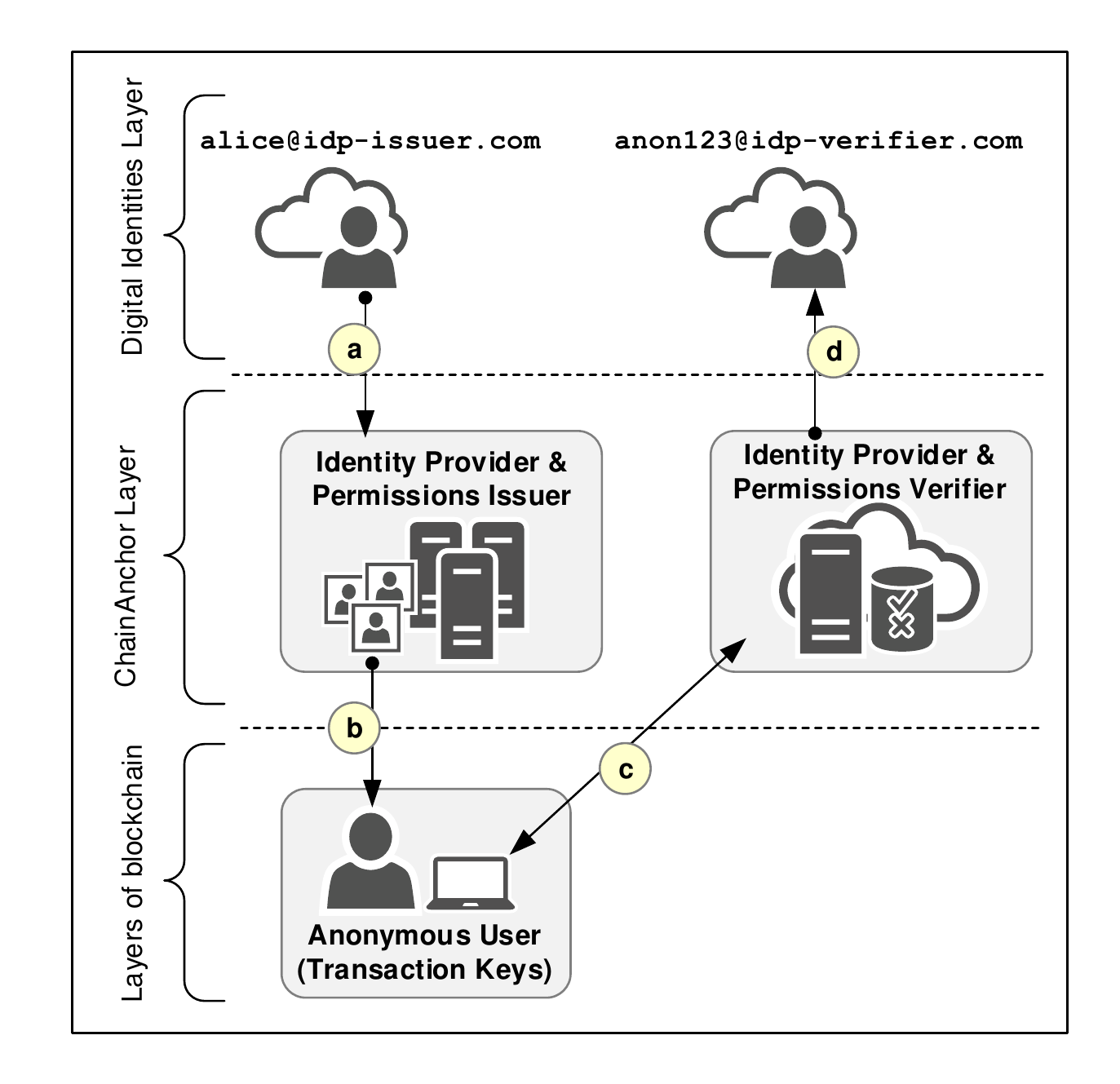}
\caption{Deriving new Internet Addressable Identities}
\label{fig:chainanchor-trustrelationships}
\end{figure}

ChainAnchor allows for further anonymous and {\em addressable} identities (e.g. email-based identities) to be derived based
on the ZKP protocol that the user executed with the IdP-PV.
Figure~\ref{fig:chainanchor-trustrelationships} illustrates as follows:
\begin{enumerate}[(a)]

\item	{\em User's known Internet identity and the IdP-PI}:
When a user seeks to participate within
a permissioned group, he or she must be approved by the group-owner (creator).
This is a normal requirement, particularly for ledgers shared by business participants.
In doing so the user must use a real world identity.
This implies that some degree of {\em personally-identifying information} (PII)
must be disclosed from the user to the group-owner (and possibly also the IdP-PI entity that is
hosting the permissioned group for the owner).
In Figure~\ref{fig:chainanchor-trustrelationships}(a)
we show this real-world identity as Alice's Internet identity
denoted by {\tt alice@idp-issuer.com}.
We assume here the IdP-PI knows all attributes relating to Alice.

\item	{\em Regaining anonymity on the permissioned blockchain}:
After the User has been approved to join the permissioned group
for the blockchain,
the user is able to regain anonymity by using
cryptographic {\em blinding} function in Step~3 of the protocol.
This blinding function prevents the IdP-PI entity
from knowing that the User-Member Private Key that was generated
by the person (Alice) whose Internet identity is {\tt alice@idp-issuer.com}
in Figure~\ref{fig:chainanchor-trustrelationships}.

\item	{\em Anonymity of user's transaction public-key}:
The anonymity of the User (obtained from the blinding function)
is further protected by the zero knowledge proof protocol
that is executed between the User and the IdP-PV
(Figure~\ref{fig:chainanchor-trustrelationships}(b)).
The IdP-PV entity has no way
of correlating between the User's transaction public-key (which is self-generated by the User)
and the Internet identity {\tt alice@idp-issuer.com}
employed by the User (Figure~\ref{fig:chainanchor-trustrelationships}(a)) to initially engage
the Permissions Issuer entity.

\item	{\em Derived anonymous Internet identity for the key-holder}:
The cryptographically anonymous relationship between the key-holder (i.e. our User)
and the IdP-PV in Figure~\ref{fig:chainanchor-trustrelationships}(c) 
lends to the possible creation by the IdP-PV
of a new anonymous Internet identity for the User,
shown as {\tt anon123@idp-verifier.com} in 
Figure~\ref{fig:chainanchor-trustrelationships}(d).

\end{enumerate}
Thus, in Figure~\ref{fig:chainanchor-trustrelationships}(d)
the IdP-PV can become an Identity Provider
and can issue a new identity {\tt anon123@idp-verifier.com}
for the anonymous user (Alice).
Furthermore, IdP-PV can bind (e.g. in an X509 certificate)
this new identity
with the transaction public-key $K_{trans}$
whose private-key ($K_{trans}^{-1}$)
is known only to the anonymous user (Alice).
We believe this approach provides a more scalable solution
than PGP.

\section{Conclusions \& Further Work}
\label{sec:furtherwork}

In this paper we have addressed the issue of identity and access control
within shared permissioned blockchains.
We proposed the ChainAchor system that provides
anonymous but verifiable identities for entities on the blockchain.

ChainAnchor allows participants in a shared permissioned blockchain
to maintain their privacy by allowing them to use
a {\em verifiable anonymous identity} when transacting on the blockchain.
The anonymous (but verifiable) identities make-up the ``group''
of entities allowed to access the shared permissioned blockchain.

In ChainAnchor the term ``consensus'' includes the notion of {\em membership} to a permissioned (private) blockchain.
An entity who has proved their membership using the zero knowledge proof protocol
can  ``register'' their self-asserted transaction
public-key (for transacting on the blockchain).
The consensus nodes collectively enforce access control
to the blockchain by only processing transactions from the members'
transaction public-keys.
Transactions belonging to unknown public-keys (i.e. non-members) are simply ignored or dropped.

A member of a shared blockchain can execute the 
zero knowledge proof protocol as many times as they desire,
each time registering a different self-asserted transaction public-key.
No one on the blockchain (including the owner of the blockchain)
has the ability to link these keys to the single owner.
This allows a member to optionally disclose their identity
(e.g. when challenged in a regulatory context)
without endangering their other transactions public-keys.
~~\\


\appendices 


\section{Summary of EPID}
ChainAnchor employs the EPID scheme~\cite{BrickellLi2012}
due to a number of advantages of the scheme.
EPID is an extension of the {\em Direct Anonymous Attestation} protocol (DAA)~\cite{Brickell2004}
for user privacy in the TPMv1.2 hardware~\cite{TPM2003}.
The EPID protocol  can be deployed without {\em Trusted Platform Module} (TPM) hardware,
with the option to add and enable a tamper-resistant TPM at a later stage.
This option may be attractive to service providers
who may wish to deploy TPM-based infrastructure in a phased approach (see~\cite{HardjonoTPM2004,TCG-IWG-2005-Thomas-Ned}).
When a TPM hardware is deployed,
it can be used to provide protected storage
for the various keys used in the ChainAnchor system.

EPID is not the only anonymous identity protocol available today.
The work of Brickell {et al.}~\cite{Brickell2004} introduced the first RSA-based DAA protocol in 2004.
A related anonymity protocol called {\em Idemix}~\cite{Camenisch2002} employs
the same RSA-based anonymous credential scheme as the DAA protocol.
However, Idemix cannot be used with the {TPMv1.2} hardware (or the new {TPMv2.0} hardware).
Another related protocol called {\em U-Prove}~\cite{UProve2014} can be integrated
into the {TPM2.0} hardware (see~\cite{Chen2013}).
However, the U-Prove protocol has
the drawback that it is not multi-show unlinkable~\cite{Chen2015},
which means that a U-Prove token may only be used once in order to remain unlinkable.

In the following we summarize the RSA-based EPID scheme
as defined in~\cite{BrickellLi2012}.

\subsection{Issuer Setup}

In order to create a group membership verification instance,
the Issuer must choose a {\em Group Public Key) and compute
a corresponding {\em Group-Issuing Private Key).

For the Group-Issuing Private Key the Issuer chooses
an RSA modulus $N = p_N q_N$ where
$p_N = 2 {p\prime}_N + 1$
and
$q_N = 2 {q\prime}_N + 1$
and where
$p_N$,
$p_N$,
${p\prime}_N$
and
${q\prime}_N$
are all prime.

The Group Public Key for the particular group instance will be:
\begin{equation}
\label{eq:group-public-key}
(N, g\prime, g, h, R, S, Z, p, q, u)
\end{equation}

The Group Issuing Private Key (corresponding to the Group Public Key) is denoted as:
\begin{equation}
\label{eq:group-issuing-private-key}
({p\prime}_N, {q\prime}_N)
\end{equation}
which the Issuer keeps secret).

In order to communicate securely with a User,
the Issuer is assumed to possess the usual
long-term public key pair
denoted as $(K_I, {K_I}^{-1})$,
where $K_I$ is publicly know in the ecosystem.

Any User who has a copy of the Group Public Key
can verify this public key by checking the following:

\begin{itemize}
\item	Verify the proof that $g, h \in \langle g\prime \rangle$
and  $R, S, Z \in \langle h \rangle$.

\item	Check whether $p$ and $q$ are primes, and check that
$q \mid (p - 1)$, $q \not| \frac{(p - 1)}{q}$ 
and ${u^q} \equiv 1 \pmod p$

\item	Check whether all group public key parameters have the required length.

\end{itemize}

\subsection{Join Protocol: User and Issuer}

In the join protocol, a given User seeks to send to the Issuer
the pair $(K, U)$ which are computed as follows.
\begin{itemize}

\item	The User chooses a secret $f$ and seeks to convey to the Issuer
a {\em commitment} to $f$ in the form of the value $U$.

\item	The value $U$ is computed as 
\begin{equation}
\label{eq:user-private-U}
U = R^{f} S^{v\prime}
\end{equation}
where $v\prime$ is chosen randomly by the User for the purpose
of {\em blinding} the chosen $f$.

\item	Next the User computes
\begin{equation}
\label{eq:user-private-f}
K = {B_I}^{f} \pmod p
\end{equation}
where $B_I$ is derived from the {\em basename} of the Issuer (denoted as ${bsn}_I$).

\end{itemize}
The goal here is for the User to send $(K, U)$ to the Issuer
and to convince the Issuer that the values $K$ and $U$ are formed correctly.

In the above Equation~\ref{eq:user-private-f},
a User chooses a base value $B$ and then uses it to compute $K$.
The purpose of the $(B, K)$ pair is for a revocation check. 
We refer to $B$ the {\em base} and $K$ as the {\em pseudonym}. 
To sign an EPID-signature, the User needs to both prove that it has
a valid membership credential and also prove that it had constructed the $(B, K)$ pair
correctly, all in zero-knowledge.
In EPID and DAA, there are two (2) options to compute the base $B$: 
\begin{itemize}

\item	{\em Random base}: Here $B$ is chosen randomly each time by the User.
A different base used every time the EPID-signature is performed.
Under the decisional Diffie-Hellman assumption, 
no Verifier entity will be able to link two EPID-signatures 
using the $(B, K)$ pairs in the signatures.

\item	{\em Named base}: Here $B$ is derived from the Verifier's basename.
That is, a deterministic function of the name of the verifier is used as a base.
For example, $B$ could be a hash of the Verifier's basename.
In this named-base option, the value
$K$ becomes a ``pseudonym'' of the User with regard to the Verifier's basename.
The User will always use the same $K$ in the EPID-signature to the Verifier.
\end{itemize}

\subsection{Issuer generates User's Membership Private Key}

In response, the Issuer performs the following steps:
\begin{itemize}

\item	The Issuer chooses a random integer $v\prime\prime$ and a random prime $e$.

\item	The Issuer computes $A$ such that 
$${ {A^e} U {S^{v\prime\prime}} } \equiv Z \pmod p$$

\item	The Issuer sends the User the values 
\begin{equation}
\label{eq:issuer-keying-material}
(A, e, {v\prime\prime}) 
\end{equation}

\end{itemize}

Note that the CL-signature~\cite{CamenischLysyanskaya2002} on the value $f$ 
is $(A, e, v := v\prime + v\prime\prime)$.
As such, the User then sets his/her Membership Private Key as:
\begin{equation}
\label{eq:membership-private-key}
(A, e, f, v)
\end{equation}
where $v := v\prime + v\prime\prime$.  
Recall that $f$ is the secret chosen by the User at the start of the Join protocol.

\subsection{User proving valid membership}

When a User seeks to prove that he or she is a group member,
the User interacts with the Verifier entity.
This is performed using the Camenisch-Lysyanskaya (CL) signature~\cite{CamenischLysyanskaya2002} on some value $f$.

This can be done using a zero-knowledge proof of knowledge
of the values $f$, $A$, $e$, and $v$ such that
\begin{equation}
\label{eq:proving-membership}
{ {A^e} {R^f} {S^v}} \equiv Z \pmod N
\end{equation}

The User also needs to perform the following:
\begin{itemize}

\item	The User computes $K = {B}^{f} \pmod p$
where $B$ is a random base (chosen by the User).

\item	The User reveals $B$ and $K$ to the Verifier.

\item	The User proves to the Verifier that the value $\log _B K$ is the same 
as in his/her private key (see Equation~\ref{eq:user-private-f}).

\end{itemize}

In proving membership to the Verifier,
the User as the prover needs to send
the Verifier the value
\begin{equation}
\label{eq:three-sigma}
\sigma = ({\sigma}_1, {\sigma}_2, {\sigma}_3)
\end{equation}
where each of the values are as follows:
\begin{itemize}

\item	${\sigma}_1$: The value ${\sigma}_1$ is a ``signature of knowledge''
regarding the User's commitment to the User's private key
and that $K$ was computed using the User's secret value $f$.

\item	${\sigma}_2$: The value ${\sigma}_2$  is a ``signature of knowledge''
that the User's private key has not been revoked by the Verifier
(i.e. not present in the signature revocation list sig-RL).

\item	${\sigma}_3$: The value ${\sigma}_3$  is a ``signature of knowledge''
that the User's private key has not been revoked by the Issuer
(i.e. not present in the issuer revocation list Issuer-RL).

\end{itemize}






%





\end{document}